%% file: text.tex
\documentclass[amsmath,amssymb,superscriptaddress,twocolumn,pra]{revtex4}
\usepackage{epsfig,appendix}

\begin{document}

\title{Parametric down-conversion from a wave-equations approach:
geometry and absolute brightness.}
\author{Morgan W. Mitchell}
\affiliation{ICFO-Institut de Ciencies Fotoniques, Mediterranean
Technology Park, 08860 Castelldefels (Barcelona), Spain}
\date{26 September 2008}

\begin{abstract}
Using the approach of coupled wave equations, we consider
spontaneous parametric down-conversion (SPDC) in the narrow-band
regime and its relationship to classical nonlinear processes such
as sum-frequency generation.  We find simple expressions in terms
of mode overlap integrals for the absolute pair production rate
into single spatial modes, and simple relationships between the
efficiencies of the classical and quantum processes.  The results,
obtained with Green function techniques, are not specific to any
geometry or nonlinear crystal. The theory is applied to both
degenerate and non-degenerate SPDC. We also find a time-domain
expression for the correlation function between filtered signal
and idler fields.

\end{abstract}

%\frontmatter

%    \begin{center}
%    {\huge Quantum Optics for the Impatient}
%    \\ \vspace{.125in} \copyright
%    { Morgan W. Mitchell \\Spring 2007 \\~ \\ ICFO - Institut de Ciencies Fotoniques
%     \\ Castelldefels }
%    \end{center}
\parskip .125in
\parindent 0in
\maketitle
\include{NewCommands}

\newcommand{\newtext}[1]{{#1}}

\newcommand{\Green}{{\cal G}}
\newcommand{\GreenBack}{{\cal H}}
\newcommand{\Eff}{{Q}}
\newcommand{\CorrFun}{{H}}
\newcommand{\BVConst}{{\beta}}
\newcommand{\Source}{{\cal S}}

\section{Introduction}

Spontaneous parametric down-conversion (SPDC) has become a
workhorse technique for generation of photon pairs and related
states in quantum optics.  Improvements in both nonlinear
materials \cite{Fejer:1992} and down-conversion geometries have
led to a steady growth in the brightness of these sources
\cite{Kwiat:1995,Kwiat:1999,Giorgi:2003,Fiorentino:2004,
Pelton:2004,Kuklewicz:2004,Fiorentino:2005,Wolfgramm:2008}.

Applications of the bright sources include fundamental tests of
quantum mechanics, quantum communications, quantum information
processing, and quantum metrology
\cite{O'brien:2003,Groblacher:2007,Higgins:2007,Sauge:2007}.
  Although
down-conversion sources typically have bandwidths of order
$10^{11}$ Hz, for the brightest sources even the output in a
few-MHz window can be useful for experiments. This permits a new
application, the interaction of down-conversion pairs with atoms,
ions, or molecules.  Indeed, sources for this purpose have been
demonstrated \cite{Haase:2009}. Many modern applications use
single-spatial-mode collection, either for improved spatial
coherence, to take advantage of fiber-based technologies, or to
separate the source and target for experimental convenience.

Remarkably, despite the importance of bright, single-spatial-mode
sources, general methods for calculating the absolute brightness
of such a source are not found in the literature.  By absolute
brightness, we mean the number of pairs per second that are
collected, for specified beam shapes, pump power, filters and
crystal characteristics. A number of calculations study the
dependence of brightness on parameters such as beam widths or
collection angles, but these typically give only relative
brightness: the final results contain an unknown multiplicative
constant \cite{Ljunggren:2005,Kurtsiefer:2001a}. While useful for
optimizing a given source, they are less helpful when designing
new sources.  A recent paper computes the absolute brightness for
a specific geometry: gaussian beams in the thin-crystal limit
\cite{Ling:2008}.

In this paper, we calculate the absolute brightness for
narrow-band, paraxial sources.  The results are quite general, for
example they apply equally well to crystals with spatial or
temporal walk-off, for non-gaussian beams, etc. The Green-function
approach we use is well suited to describing the temporal features
of the down-conversion pairs, and we are able to predict the time
correlations in a particularly simple way.  To our knowledge, this
method of deriving the time-correlations is also novel.

Perhaps of greatest practical importance, we derive very simple
relationships between the efficiency of classical parametric
processes and their corresponding quantum parametric processes.
For example, in any given geometry the efficiency of sum-frequency
generation and spontaneous parametric down-conversion are
proportional.  This allows the use of existing classical
calculations and/or experiments with classical nonlinear optics to
predict the brightness of quantum sources.

The paper is organized as follows: In section
\ref{sec:Precedents}, we describe briefly the variety of
theoretical treatments that have been applied to parametric
down-conversion, and our reasons for making a new calculation.  In
Section \ref{sec:Formalism} we describe the formalism we use,
based on an abstract paraxial wave equation and Green function
solutions.  In Section \ref{sec:Results} we calculate the absolute
brightness and efficiencies for non-degenerate and degenerate
parametric down-conversion and the corresponding classical
processes.  In section \ref{sec:Conclusions} we summarize the
results.

\section{Background}
\label{sec:Precedents}

The characteristics of parametric down-conversion light have been
calculated in a number of different ways.  Kleinman
\cite{Kleinman:1968} used a Hamiltonian of the form \be H' =
-\frac{1}{3} \int d^3 x \, \bE \cdot \chi : \bE \ee and the Fermi
``golden rule'' to derive emission rates as a function of
frequency and angle. Zel'dovich and Klyshko \cite{Zeldovich:1969}
proposed to use a mode expansion and calculate pair rates treating
the quantum process as a classical parametric amplifier seeded by
vacuum noise . Detailed treatment along these lines is given in
\cite{Shen:1984,Klyshko:1988}.  The problem of collection into
defined spatial modes was not considered, indeed the works
emphasize that the {\em total} rate of emission is {\em
independent} of pump focusing.

After the observation of SPDC temporal correlations by Burnham and
Weinberg \cite{Burnham:1970}, Mollow \cite{Mollow:1973} described
detectable field correlation functions (coincidence distributions)
in terms of source-current correlations and Green functions of the
wave equation.  This Heisenberg-picture calculation derived
absolute brightness for multi-mode collection, e.g., for detectors
of defined area at defined positions.  It did not give brightness
for single-mode collection, nor a connection to classical
nonlinear processes.  Hong and Mandel \cite{Hong:1985} used a
mode-expansion to compute correlation functions based on the
Heisenberg-picture evolution and an interaction Hamiltonian of the
form \be H_I = \frac{1}{2} \int d^3x \, \chitwo_{ijk}
E_iE_jE_k.\ee As with Mollow's calculation, they find singles and
pair detection rates, but only for multi-mode detection
\footnote{We note that Hamiltonian-based treatments often do not
agree about the constant preceding the integral $\int d^3 x
\chitwo E^3$, not even its sign.  This question does not present a
problem for the present calculation, which does not employ a
Hamiltonian.}
.
 Ghosh, {\em et al.} \cite{Ghosh:1986} used the same Hamiltonian in
a Schr\"{o}dinger-picture description, truncating the time
evolution at first order to derive a ``two-photon wave-function."
This last method has become the most popular description of SPDC,
including work on efficient collection into single spatial
modes\cite{Ljunggren:2005,Kurtsiefer:2001a}.  Many works along
these lines are cited in reference \cite{Ljunggren:2005}.
Recently, Ling, et al.~\cite{Ling:2008} calculated the absolute
emission rate based on a similar interaction Hamiltonian and a
gaussian-beam mode expansion.  In this way, they are able to
calculate the absolute pair rate for non-degenerate SPDC in a
uniform, thin crystal into gaussian collection modes.  As
described in Section \ref{Sec:Ling}, our calculation agrees with
that of Ling et al. while also treating other crystal geometries,
general beam shapes, and degenerate SPDC.

Notable differences among the calculations include Heisenberg
{vs.} Schr\"{o}dinger picture and calculating in direct space {
vs.} inverse space via a mode expansion.  While they are of course
equivalent, Heisenberg picture calculations are easier to compare
to classical optics, while Schr\"{o}dinger picture calculations
are more similar to the state representations in quantum
information. As our goal is in part to connect classical and
quantum efficiencies, we use the Heisenberg picture.  Also, we
note that the Schr\"{o}dinger-picture ``two-photon wave-function''
 has a particular pathology: the first-order treatment of
time evolution means the Schr\"{o}dinger picture state is not
normalized and never contains more than two down-conversion
photons.  While this is not a problem for calculation of relative
brightness or pair distributions\cite{Valencia:2007,Mosley:2008},
it does prevent calculation of absolute brightness. The choice of
inverse { vs.} real space calculation is also one of convenience:
for large angles in birefringent media or detection in momentum
space, plane waves are the ``natural'' basis for the calculation.
However, most bright sources use paraxial geometries and
collection into defined spatial modes, e.g., the gaussian modes of
optical fibers. In these situations, the advantages of a mode
expansion disappear, while the local nature of the $\chitwo$
interaction makes real-space more ``natural.''  Thus we opt for a
real-space calculation.

Our treatment of SPDC is based on coupled wave equations, a
standard approach for multi-wave mixing in non-linear optics
\cite{Boyd:2008}.  The calculations are done in the Heisenberg
picture, so that the evolution of the quantum fields is exactly
parallel to that of the classical fields described by nonlinear
optics.   This allows the re-use of well-known classical
calculations such as those by Boyd and Kleinman \cite{Boyd:1968}.
As in the approach of Mollow, we use Green functions to describe
the propagation, and find results that are not specific to any
particular crystal or beam geometry. Unlike Mollow's calculation,
we work with a paraxial wave equation (PWE). This allows us to
simply relate the classical and quantum processes through
momentum-reversal, which takes the form of complex conjugation in
the PWE.

We focus on narrow-band parametric down-conversion, for which the
results are particularly simple. By narrow-band, we mean that the
bandwidths of the pump and of the collected light are much less
than the bandwidth of the SPDC process, as set by the
phase-matching conditions.  This includes recent experiments with
very narrow filters \cite{Haase:2009}, but also a common
configuration in SPDC, in which the down-conversion bandwidth is
$\sim 10$ nm while the filter bandwidths are $< 1$ nm.

\section{Formalism}
\label{sec:Formalism}

\subsection{description of propagation}

We are interested in the envelopes $\tbE_\pm$ for forward- and
backward-directed of parts of the quantum field $E^{(+)}(t,\bx) =
(\tbE_{+}\exp[+ i k z] + \tbE_{-}\exp[- i k z ] )\exp[- i \omega
t]$ where $k$ is the average wave-number and $\omega$ is the
carrier frequency. These propagate according to a paraxial wave
equation \bea \label{Eq:DiffEq}{\cal D}_{\pm} \tbE_{\pm} &=&
\Source_\pm,\eea where $\cD_{\pm}$ is a differential operator and
$ \Source_\pm$ is a source term (later due to a $\chitwo$
non-linearity).

The formal (retarded) solution to equation (\ref{Eq:DiffEq}) is
\be \tbE_{\pm}(x) = \tbE_{0\pm}(x)+\int d^4 x' \Green_\pm(x;x')
\Source_\pm(x') \ee where $x$ is the four-vector $(t,\bx)$,
$\tbE_{0\pm}(x)$ is a solution to the source-free ($\Source = 0$)
equation, and $\Green_\pm$ are the time-forward Green functions,
defined by \bea \cD_\pm \Green_\pm(x;x') &=& \delta^4(x-x') \nn
\Green_\pm(x;x') & = & 0 ~~~~~ t < t'. \eea

\newtext{For illustration, we consider the paraxial wave equation (PWE),
for which
%\be \label{Eq:PWE} \cD_\pm \equiv \partial2_x +
%\partial2_y \pm i {\beta_x}\partial_x + i {\beta_y}\partial_y  \pm
% 2ik(\partial_z \pm v_{g}^{-1} \partial_t )
%\ee
%\be \label{Eq:PWE} \cD_\pm \equiv \nabla_T^2  \pm 2ik(\partial_z
%\pm v_{g}^{-1} \partial_t ) \ee
\bea \label{Eq:PWE} \cD_\pm &\equiv& \nabla_T^2  \pm
2ik(\partial_z \pm v_{g}^{-1} \partial_t ) \\
\Source_\pm &=& \frac{\omega^2}{c^2 \varepsilon_0} {\cal P}^{(\rm
 NL)}_\pm.
\eea
 Here $\nabla_T^2$ is the
transverse Laplacian, $k = n(\omega) \omega /c$ is the
wave-number, $v_g \equiv
\partial \omega /
\partial k_z$ is the group velocity, and $\cal P^{(\rm NL)}$ is the
envelope for the nonlinear polarization.}

 We note that $\cD_\pm$
is invariant under translations of $x$, and that time reversal
$t\rightarrow -t$ is equivalent to direction-reversal and complex
conjugation, i.e., $\cD_\pm \rightarrow \cD_\mp^*$ .
%$\cD_\pm\tbE_\pm = \Source_\pm \rightarrow \cD_\mp^*\tbE^*_\mp = \Source^*_\mp $
The results we obtain will be valid for any equation obeying these
symmetries.  In particular, the results will also apply to
propagation with dispersion and/or spatial walk-off, which can be
included by adding other time and/or spatial derivatives to $\cD$.

From the symmetries of $\cD_\pm$, it follows that the Green
functions depend only on the difference $x-x'$, and that
$\Green_+(t,\bx;t',\bx') = \Green^*_-(t,\bx',t',\bx)$. Also, the
time-backward (or ``advanced'') Green functions $\GreenBack_\pm$,
defined by \bea \cD_\pm \GreenBack_\pm(x;x') &=& \delta^4(x-x')
\nn \GreenBack_\pm(x;x') & = & 0 ~~~~~ t > t'\eea obey
$\GreenBack_\pm(x;x') = \Green_\pm^*(x',x)$.

%$\Green_+(x;x') = g(t-t',\bx-\bx')$, then $\Green_-(x;x') =
%g^*(t-t',\bx'-\bx)$

%We will also need the time-backward (or ``advanced'') Green
%functions $\GreenBack_\pm$, defined by \bea \cD_\pm
%\GreenBack_\pm(x;x') &=& \delta^4(x-x') \nn \GreenBack_\pm(x;x') & = &
%0 ~~~~~ t > t'.\eea  As shown in the Appendix, $\GreenBack_\pm(x;x') =
%\Green_\pm^*(x',x)$ and
%In fact, as shown in Appendix
%\ref{App:TimeReversal}, this is a property of any equation showing
%time-reversal symmetry, and thus our results apply to a broad
%class of propagation situations, including propagation with
%spatial walk-off and group-velocity dispersion.

\subsection{boundary and initial value problems}

If the value of the field is known on a plane $z=z_{\rm src}$, the
field downstream of that plane is \be
\label{Eq:BoundaryValueProblem} \tbE_\pm(x) = \BVConst_z \int
d^4x' \Green_\pm(x;x') \tbE_{\pm}(x') \delta(z'-z_{\rm src})\ee
where $\BVConst_z \equiv \pm 2ik$. Similarly, if the field is
known at an initial time $t=t_0$, the field later is
\be\label{Eq:InitialValueProblem} \tbE_\pm(x) = \BVConst_t \int
d^4x' \Green_\pm(x;x')\tbE_\pm(x')\delta(t'-t_0)\ee where
$\BVConst_t = 2ik/v_g$.  Similar relationships hold for the
advanced Green functions.  If the field is known in some plane
$z=z_0$ downstream, then \bea \label{Eq:BVProblem2} \tbE_{\pm}(x)
&=& \BVConst_{z}^* \int d^4 x'\,
\GreenBack_\pm(x;x')\tbE_{\pm}(x')\delta(z'-z_0)
 \nne \BVConst_{z}^* \int d^4 x'\,
\tbE_{\pm}(x')\delta(z'-z_0)\Green^*_\pm(x';x)  \eea while if the
field is known at some time $t_f$ in the future, \bea
\label{Eq:BackFromTheFuture} \tbE_{\pm}(x) &=& \BVConst_{t}^* \int
d^4 x'\, \GreenBack_\pm(x;x')\tbE_{\pm}(x')\delta(t'-t_f)
 \nne \BVConst_{t}^* \int d^4 x'\,
\tbE_{\pm}(x')\delta(t'-t_f)\Green^*_\pm(x';x)  \eea

%or $\Green_-(x;x') = \Green_+^*(t,\brho,z',t',\brho',z)$.

\subsection{quantization}

The field envelopes are operators which obey the equal-time
commutation relation \be \label{Eq:EqualTimeCommutator}
[\tbE(\bx,t),\tbE^\dagger(\bx',t)] = A_{\gamma}^2
\delta^3(\bx'-\bx)\ee where $A_{\gamma} \equiv \sqrt{\hbar \omega
/2 n n_g \varepsilon_0}$ is a photon units scaling factor and $n_g
\equiv c/v_g$ is the group index. For narrow-band fields,
$A_{\gamma}^{-2}\expect{\tbE^\dagger\tbE} $ describes a photon
number density, and $v_g A_{\gamma}^{-2}\expect{\tbE^\dagger\tbE}$
and $v_{gs}v_{gi} A_{\gamma i}^{-2}A_{\gamma s}^{-2}
\expect{\tbE_s^\dagger\tbE_i^\dagger\tbE_i\tbE_s}$ describe single
and pair fluxes.  We find the unequal-time commutation relation
from equation (\ref{Eq:InitialValueProblem}) \be\left.
[\tbE(x),\tbE^\dagger(x')]\right._{t>t'} = \BVConst_t A_{\gamma}^2
\Green(x;x')\ee so that $\bra{0}{\tbE(x)\tbE^\dagger(x')}\ket{0} =
\BVConst_t A_{\gamma}^2 \Green(x;x')$ for $t>t'$.  For the PWE,
$A_{\gamma}^{-2} v_g = 2 n c \varepsilon_0/\hbar \omega$ and
$\BVConst_t A_{\gamma}^2 = i \hbar \omega^2 /c^2\varepsilon_0$.

To calculate singles rates, we will need to evaluate expressions
of the form $\expect{\tbE\tbE^\dagger}$.  For this, a useful
expression is derived in the Appendix:  Equation
(\ref{Eq:AlternatePropagator})
 \bea
\expect{\tbE_{}(x)\tbE_{}^\dagger(x') }&=&
  \frac{2\hbar n \omega^3 }{c^3 \varepsilon_0}
\int d^4 x'' \delta(z''-z_0) \nnt  \Green^*(x'';x) \Green(x'';x').
\eea Here $z_0$ is any plane down-stream of $x$ and $x'$.

\subsection{single spatial modes}

A single spatial mode $M_\pm(\bx)$ is a time-independent solution
to the source-free wave equation $\cD_\pm M_\pm(\bx)=0$.
$M_\pm^*(\bx)$ is the corresponding momentum-reversed solution
$\cD_\mp M_\pm^*(\bx)=0$. We assume the normalization $\int d^3x
|M_\pm(\bx)|^2 \delta(z) = 1$. For single-mode collection, it will
be convenient to define the projection of a field $\cE(x)$ onto
the mode $M$ as \be \cE_M(t) \equiv \int d^3x
M^*(\bx)\delta(z-z_0) \cE(x)\ee  (here and below, the $+/-$
propagation direction is the same for $\cE,M$).  Here $z_0$ is
some plane of interest, and $\cE_M(t)$ describes the magnitude of
the field component in this plane.  Similarly, if the envelope is
constant, the field distribution is \be \cE(x) = \cE_M(t) M(\bx)
.\ee  The optical power is (MKS units) $P_M(t) = 2 n
c\varepsilon_0 \int d^3x |\tbE(t,x)|^2\delta(z-z_0) = 2 n
c\varepsilon_0 |\tbE_M(t)|^2$.

%For considering collection of radiated fields into single modes,
%it is very convenient to note that Equation
%(\ref{Eq:BoundaryValueProblem}) and the time-translation symmetry
%of $\Green$ imply \bea \label{Eq:PropToOverlap1} \Green_M(t;x')
%&\equiv & \int d^3x d^4x' M^*(\bx)\delta(z-z_0) \Green(x;x') \nne
%\frac{1}{\BVConst_z} \int d^3x' M^*(\bx').\eea

Given an upstream source $\Source(x)$, the  $M$ component of the
generated field is \bea \cE_M(t) &=& \int d^3x d^4x'
M^*(\bx)\delta(z-z_0) \nnt \Green(x;x') \Source(x').\eea
%\bea
%\cE_M(t) &=& \cE_{0M}(t)+\int d^3x d^4x' M^*(\bx)\delta(z-z_0)
%\nnt \Green(x;x') \Source(x').\eea
If the source is time-independent, then Equation
(\ref{Eq:BVProblem2}) and the time-translation symmetry of
$\Green$ imply \be \cE_M(t) = \frac{1}{\BVConst_z} \int d^3x'
M^*(\bx') \Source(x').\ee
%\be \cE_M(t) = \cE_{0M}(t) + \frac{1}{\BVConst_z} \int d^3x'
%M^*(\bx') \Source(x').\ee

Similarly, if a product $\cE_1(x_1)\cE_2(x_2)$ is given by a
constant pair source $\Source^{(2)}(x)$ as \bea
\cE_1(x_1)\cE_2(x_2) &=& \int d^4x'
\Green_1(x_1;x')\Green_2(x_2;x') \nnt \Source^{(2)}(x').\eea then
the time-integrated mode-projected component is \bea \int dt_1
\cE_{1M_1}(t_1)\cE_{2M_2}(t_2) &=&
\frac{1}{\BVConst_{1z}\BVConst_{2z}} \int d^3x' M_1^*(\bx')\nnt
M_2^*(\bx') \Source^{(2)}(x').\eea

\subsection{Coupled wave equations}

We now introduce a $\chitwo$ nonlinearity, which produces a
nonlinear polarization that appears as a source term in the
propagation equations.  We consider three fields, ``signal,''
``idler'' and ``pump'' with carrier frequencies $\omega_s,
\omega_i,\omega_p$ and wave-numbers $k_s,k_i,k_p$, respectively.
The respective field envelopes $\tbE_s,\tbE_i,\tbE_p$ evolve
according to \bea {\cal D}_p \tbE_p &=& \omega_p^2 g \tbE_s \tbE_i
\exp[i \Delta k z]\nn {\cal D}_s \tbE_s &=& \omega_s^2 g \tbE_p
\tbE_i^\dagger \exp[-i \Delta k z] \nn {\cal D}_i \tbE_i &=&
\omega_i^2 g \tbE_p \tbE_s^\dagger \exp[-i \Delta k z]\eea where
$g = -4m(\bx)d /c^2 $, $d$ is the effective nonlinearity, equal to
half the relevant projection of $\chitwo$,  and $\Delta k \equiv
k_p-k_s-k_i$ is the wave-number mismatch.  The dimensionless
function $m(\bx)$ describes the distribution of $\chitwo$.  For
example in a periodically-poled material it alternates between
$\pm 1$.   We can take $\Delta k=0$ without loss of generality, as
the phase oscillation can be incorporated directly in the
envelopes. The propagation directions ($\pm$) will be omitted
unless needed for clarity. Note that for transparent materials
$\chitwo$ is real, and $\chitwo(\omega_p;\omega_s+\omega_i)=
\chitwo(\omega_s;\omega_p-\omega_i)=
\chitwo(\omega_i;\omega_p-\omega_s)$.

% and $\cD_m$ is the differential operator for the
%paraxial wave equation
%%\be  \nabla2 - \frac{n2}{c^2}
%%\partial2_t \ee or
%%\be \label{Eq:PWE} \cD_m \equiv \frac{-i
%%v_{gm}}{2k_m}(\partial2_x +
%%\partial2_y) +
%%  v_{gm}\partial_z + \partial_t.
%%\ee
%\be \label{Eq:PWE} \cD_m \equiv \nabla_T2 +
% 2ik_m(\partial_z + v_{gm}^{-1} \partial_t )
%\ee
% The subscripts $m={p,s,i}$ indicate pump, signal and idler
%respectively.  We express the solution in terms of the forward
%Green function $\Green_m$, defined by \bea \cD_m \Green(x;x') &=&
%\delta4(x-x') \nn \Green(x;x') & = & 0 ~~~~~ t < t' \eea where $x$ is
%the four-vector $(\bx,t)$.

First-order perturbation theory is sufficient to describe
situations in which pairs are produced.  For example, if
$\tbE_{0s},\tbE_{0i},\tbE_{0p}$ are source-free solutions,
 then \bea \label{Eq:EsToFirstOrder} \tbE_s & = & \tbE_{0s} + \omega_s^2
\int d^4x' \Green_{s}(x;x') \nnt g(x')
\tbE_{0p}(x')\tbE_{0i}^\dagger(x') + O(g^2).\eea
%If $\tbE_{0m}$ are solutions for $g=0$, we have first-order
%solutions \bea \tbE_p & = & \tbE_{0p} + \omega_p^2 \int d^4x'
%\Green_{p}(x;x') \nnt g(x') \tbE_{0s}(x')\tbE_{0i}(x') \nn \tbE_s
%& = & \tbE_{0s} + \omega_s^2 \int d^4x' \Green_{s}(x;x')  \nnt
%g(x') \tbE_{0p}(x')\tbE_{0i}^\dagger(x')  \nn \tbE_i & = &
%\tbE_{0s} + \omega_i^2 \int d^4x' \Green_{i}(x;x')  \nnt g(x')
%\tbE_{0p}(x')\tbE_{0s}^\dagger(x').
%  \eea
and similar expressions for $\cE_i,\cE_p$ are sufficient to give
the lowest-order contribution to the pair-detection rate
$W^{(2)}\propto \expect{\tbE^\dagger_{s} \tbE^\dagger_{i}
\tbE_{i}\tbE_{s}}$. Higher-order expansions would be necessary for
double-pair production, etc.

\subsection{narrow-band frequency filters}

In most down-conversion experiments, some sort of frequency filter
is used.
%In the frequency domain, a linear, stationary filter is
%defined by its transfer function $T(\Omega)$, \bea
%\tbE^{(F)}(\Omega) &\equiv& \tbE_{\rm out}(\Omega) \nne
%T(\Omega)\tbE_{\rm in}(\Omega)+R(\Omega)\tbE_{\rm
%res}(\Omega).\eea Here  $|R(\Omega)|^2 + |T(\Omega)|^2=1$, where
%$R(\Omega)$ is the coupling to a reservoir field $\tbE_{\rm res}$.
%We assume this reservoir is in the vacuum state, and thus will not
%produce any detections and can be safely ignored. $T(\Omega)$ is
%the Fourier transform of the impulse response $F(\tau)$.
Assuming this filter is linear and stationary, the field reaching
the detector is \be \tbE^{(F)}(t) = \int dt' F(t-t')\tbE(t') +
G(t-t')\tbE_{\rm res}(t').\ee  Here $\tbE_{\rm res}$ is a
reservoir field required to maintain the field commutation
relations.  Assuming the reservoir is in the vacuum state, it will
not produce detections and can be ignored. Defining
%\footnote{Note that we have two very
%different things called "$G$."  The first is the Green function
%$G(x;x')$ named after the 19th century mathematician and physicist
%George Green, and the second is the intensity correlation function
%$\CorrFun$, introduced into quantum optics by Roy Glauber.}
$\CorrFun_F(t_i,t_s) \equiv \expect{ \tbE^{(F_i)}_i(t_i)
\tbE^{(F_s)}_s(t_s) },$ the fields that leave the filter obey \bea
\CorrFun_F(t_i,t_s) &=& \int dt' dt'' F_i(t_i-t') F_s(t_s-t'')
\nnt \expect{ \tbE_i(t') \tbE_s(t'') }.\eea  In the narrowband
case, i.e., when the correlation time between signal and idler is
much less than the time-scale of the impulse response functions,
we can take $\expect{ \tbE_i(t') \tbE_s(t'') } \approx {\cal A}
\delta(t'-t'')$ where the constant ${\cal A} \equiv \int dt_i
\expect{\tbE_i(t_i)\tbE_s(t_s)}$. We find \bea \CorrFun_F(t_i,t_s)
&\approx& {\cal A}\int dt' F_i(t_i-t') F_s(t_s-t') \nnequiv {\cal
A} f(t_s-t_i) .\eea  With this, we see that the flux of pairs is
\bea W^{(2)}(t_s-t_i) &=&
%A_{\gamma i}^{-2}A_{\gamma
%s}^{-2}v_{gs}v_{gi} |{\cal A}f(t_s-t_i)|^2 \nne
\frac{4 n_s n_i c^2 \varepsilon_0^2}{\hbar^2 \omega_s \omega_i }
|{\cal A}f(t_s-t_i)|^2 \eea with a total coincidence rate of \bea
W^{(2)} &=& \int dt_i W^{(2)}(t_s-t_i) \nne \frac{n_s n_i c^2
\varepsilon_0^2}{\hbar^2 \omega_s \omega_i } |{\cal A}|^2 \int
dt_i |2 f(t_s-t_i)|^2 \nnequiv \frac{n_s n_i c^2
\varepsilon_0^2}{\hbar^2 \omega_s \omega_i } |{\cal A}|^2
\Gamma_{\rm eff}.\eea  We note that \be \label{Eq:GammaArea}
\Gamma_{\rm eff} = \frac{2}{\pi} \int d\Omega
T_s(\Omega)T_i(-\Omega)\ee where $T_{s,i}(\Omega) \equiv |\int dt
\exp[i \Omega t] F_{s,i}(t)|^2$ are the signal and idler filter
transmission spectra, respectively. For this reason we refer to
$\Gamma_{\rm eff}$ as the effective line-width (in angular
frequency) for the combined filters. Also important will be the
singles rate \bea \label{Eq:SinglesRateGeneral}W^{(1)} &=&
A_{\gamma s}^{-2}v_{gs}
\expect{[\tbE_s^{(F_S)}(t_s)]^\dagger\tbE_s^{(F_S)}(t_s)}\nne
\frac{2 n_s c \varepsilon_0 }{\hbar \omega_s} \int dt' dt''
F_s^*(t_s-t') F_s(t_s-t'') \nnt \expect{ \tbE_s^\dagger(t')
\tbE_s(t'')} \nnapprox \frac{2 n_s c \varepsilon_0 }{\hbar
\omega_s} {\cal C} \int dt'  |F_s(t_s-t')|^2  \nnequiv \frac{n_s c
\varepsilon_0 }{2 \hbar \omega_s} {\cal C} \Gamma_{{\rm eff},s}
\eea where ${\cal C} \equiv \int dt' \expect{ \tbE_s^\dagger(t')
\tbE_s(t'')}$. $\Gamma_{{\rm eff},s}$ is the effective line-width
for the signal filter.

\section{Results}
\label{sec:Results}

With the calculational tools described above, we now demonstrate
the central results of this paper.  We first express the
efficiency of continuous-wave sum-frequency generation (SFG) in
terms of a mode-overlap integral.  This effectively reduces the
non-linear optical problem to three uncoupled propagation
problems.  We then show that the efficiency of parametric
down-conversion in the same medium is proportional to the SFG
efficiency, for modes with the same shapes but opposite
propagation direction.  The constant of proportionality is found,
allowing calculations of absolute efficiency based either on
material properties such as $\chitwo$ or measured SHG
efficiencies.  Similarly, the singles production efficiency is
related to difference-frequency generation (DFG) and the
collection efficiency is calculated.  The same quantities for the
degenerate case are also found.

\subsection{sum-frequency generation}

We consider first the process of SFG, for un-depleted signal and
idler and no input pump.  Signal and idler are constant and come
from single-modes, \bea \label{Eq:SingleModeSFGBrightness}
\tbE_{M_p}(t_p) & = & -\tbE_{M_i}\tbE_{M_s} \frac{4\omega_p^2 d
}{c^2\BVConst_{z,p}} \nnt \int d^3x' M_p^*(x')m(x') M_i(x')M_s(x')
\nnequiv -\tbE_{M_i}\tbE_{M_s}\frac{4\omega_p^2
d}{c^2\BVConst_{z,p}} I_{SFG}.\eea

 The conversion efficiency is \bea \label{Eq:SingleModeSFGEff}
\Eff_{SFG}^{} &\equiv& \frac{P_{M_p}}{P_{M_s}P_{M_i}} =
\frac{8\omega_p^4 n_p d^2|I_{SFG}|^2}{ n_s n_i c^5 \varepsilon_0
|\BVConst_{z,p}|^2} \nne \frac{2 \omega_p^2 d^2}{c^3 \varepsilon_0
n_p n_s n_i }|I_{SFG}|^2\eea

%Assuming that signal and idler come from single-modes, \bea
%\tbE_p(x) & = & \int d^4x'dt_s dt_i \Green_{p}(x;x')
%g(x')\Green_{s}(x';t_s,M_s)\nnt \Green_{i}(x';t_i,M_i) \tbE_{M_i
%i}(t_i)\tbE_{M_s s}(t_s) .\eea  Finally, assuming the input fields
%are constant $\tbE(t) = \tbE_{U}$, and the sum-frequency light is
%collected into a single mode \bea
%\label{Eq:SingleModeSFGBrightness} \tbE_p(t_p) & = & \tbE_{M_i
%i}\tbE_{M_s s}\int d^4x'dt_s dt_i \Green_{p}(t_p,M_p;x')\nnt g(x')
%\Green_{s}(x';t_s,M_s)\Green_{i}(x';t_i,M_i) .\eea
%
%Noting that $\Green_p,\Green_s,\Green_i$ are functions of $t_p-t',
%t'-t_s,t'-t_i$, respectively, we can change the integration from
%$dt' dt_s dt_i$ to \mbox{$d(t_p-t')d(t'-t_s)d(t'-t_i)$} $ \equiv
%dt_1dt_2dt_3$ to get
% \bea \tbE_p(t_p) & = & \tbE_{M_i
%i}\tbE_{M_s s}\int d^3x' dt_1 dt_2 dt_3 \Green_{p}(t_1,M_p;0,x')\nnt
%g(x') \Green_{s}(0,x';t_2,M_s)\Green_{i}(0,x';t_3,M_i) \nne \tbE_{M_i
%i}\tbE_{M_s s}\int d^3x' M_p^*(x')g(x')M_s(x')M_i(x')\eea
%
% $M^*$
%appears when the mode $U$ is collected, and thus $M$ describes the
%field that reaches that mode, or more precisely, the
%back-propagated field from that mode.

The efficiency of a cw, single-mode source is thus proportional to
the spatial overlap of the pump, signal, and idler modes, weighted
by the nonlinear coupling $g$.

\subsection{non-degenerate parametric down-conversion}

Next we consider the process of parametric down-conversion.
%\bea \tbE_s & = & \tbE_{0s} + \omega_s^2 \int d^4x'
%\Green_{s}(x;x')  \nnt g(x') \tbE_{0p}(x')\tbE_{0i}^\dagger(x')
%\nn \tbE_i & = & \tbE_{0i} + \omega_i^2 \int d^4x' \Green_{i}(x;x')
%\nnt g(x') \tbE_{0p}(x')\tbE_{0s}^\dagger(x').
%  \eea
Using Equation (\ref{Eq:EsToFirstOrder}), we can calculate to
first order in $g$ the correlation function \bea
\label{Eq:NDPDCCorrFun} \expect{\tbE_{i}(x_i)\tbE_{s}(x_s)} &=&
\omega_s^2 \int d^4 x' \Green_s(x_s,x')
\nnt\expect{\tbE_{0,i}(x_i)\tbE^\dagger_{0,i}(x')}\nnt
g(x')\tbE_{0,p}(x') \nne i \frac{\hbar \omega_i^2
\omega_s^2}{c^2\varepsilon_0} \int d^4 x'
\Green_s(x_s,x')\nnt\Green_i(x_i,x') g(x')\tbE_{0,p}(x') \eea For
constant pump and single-mode collection we have \bea
\label{Eq:SingleModeSPDCA} {\cal A}_{M_i M_s} &\equiv& \int dt_s
\expect{\tbE_{M_i}(t_i)\tbE_{M_s}(t_s)} \nn &=& \frac{i
\hbar\omega_s \omega_i d  }{c^2\varepsilon_0 n_s n_i}\tbE_{M_p}
\nnt \int d^3x' M_s^*(x') M_i^*(x') m(x') M_p(x') \nnequiv \frac{i
\hbar\omega_s \omega_i d  }{ c^2\varepsilon_0 n_s n_i}\tbE_{M_p}
I_{DC} . \eea

 We note that
$I_{DC} = I_{SFG}^*$.  Also, the conjugate modes describe
backward-propagating fields, as if the source fields were sent
through the nonlinear medium in the opposite direction.    Thus if
we want to know the brightness of down-conversion when all beams
are propagating to the left, it is sufficient to calculate (or
measure) the efficiency of up-conversion when all beams are
propagating to the right.  Using equations
(\ref{Eq:SingleModeSFGEff}) and (\ref{Eq:SingleModeSPDCA}) we find
\bea \label{Eq:SingleModeSPDCBrightness} \left|{\cal A}_{M_i
M_s}\right|^2 &=& \frac{\hbar^2 \omega_i^2\omega_s^2}{4 c^2
\varepsilon_0^2 n_s n_i \omega_p^2 }P_p  \Eff_{SFG}^{} . \eea

\subsection{brightness}

We can now consider the brightness of the filtered, single-mode
source.  The rate of detection of pairs is \bea
\label{Eq:PDCPairRate}W^{(2)} &=& \frac{n_s n_i c^2
\varepsilon_0^2}{\hbar^2 \omega_s \omega_i } |{\cal A}|^2
\Gamma_{\rm eff} \nne  \Gamma_{\rm eff}
\frac{\omega_i\omega_s}{4\omega_p^2 }P_p  \Eff_{SFG}^{} \eea

 This simple expression is the first
main result: The rate of pairs is simply the joint collection
bandwidth $\Gamma_{\rm eff}$, times the ratio of frequencies,
times the pump power, times the up-conversion efficiency
$\Eff_{SFG}^{}$.  Note that the last quantity can be calculated if
the mode shapes and $\chitwo(\bx)$ are known, for example in the
paper of Boyd and Kleinman, or simulated for more complicated
situations.  Most importantly, it is directly measurable.

%As a numerical example: for a typical type-II crystal,
%$\Eff_{SFG}$ is about $2\times 10^-{3}$ W$^{-1}$, so that for 100
%mW pump power, the brightness is $1/8 \times 10^{-4}$ times the
%bandwidth $\Gamma_i$ if $\Gamma_i \ll \Gamma_s$.  !!! NOTE:  This
%estimate of $\Eff_{SFG}$ is very crude: it is based on the known
%efficiencies for type-I crystals.  Would be good to get an some
%data for type-II.

\subsection{difference-frequency generation}

We now consider the classical situation in which pump and signal
beam are injected into the crystal and idler is generated.  We
will see that this directly measurable process is related to the
singles generation rate by parametric down-conversion.  The
generated idler is \bea \tbE_i(x) & = & \omega_i^2 \int d^4x'
\Green_{i}(x;x') g(x') \tbE_{0p}(x')\tbE_{0s}^*(x').\nonumber \eea
If pump and signal are from modes $M_P,M_S$, respectively, we find
\bea \tbE_i(x) & = & -\frac{4\omega_i^2 d}{c^2}
\tbE_{M_p}(t_p)\tbE_{M_s}^*(t_s) \int d^4x' \Green_{i}(x;x') \nnt
m(x') M_{p}(x')M_s^*(x')
% \nnequiv
%\omega_i2 \tbE_{M_p}(t_p)\tbE_{M_s}^*(t_s) I_{DFG}^{(s)}
.\eea  The total power generated is $P_{i} = 2 c n_i \varepsilon_0
\int d^3 x_i  \delta(z_i-z_0)  |\tbE_i(x_i)|^2$ where $z_0$
indicates a plane downstream of the generation.  We find \bea
P_{i} & = & \frac{2\omega_i^2 d^2 }{c^3 \varepsilon_0 n_s n_i n_p
} {P_p}{P_s} \int d^3 x_i \delta(z_i-z_0)  \nnt
\left|\BVConst_{z,i}^{} \int d^4x' \Green_{i}(x_i;x') m(x')
M_{p}(x')M_s^*(x') \right|^2 \nnequiv {P_p}{P_s} \frac{
2\omega_i^2 d^2}{c^3  \varepsilon_0 n_s n_i n_p }
|I_{DFG}^{(s)}|^2 \nnequiv {P_p}{P_s} Q_{DFG}
% \nnequiv
%\omega_i2 \tbE_{M_p}(t_p)\tbE_{M_s}^*(t_s) I_{DFG}^{(s)}
. \label{Eq:PDFG}\eea

\subsection{singles rates in PDC}

We can find the rate of detection of singles in the mode $M_S$ by
equation (\ref{Eq:SinglesRateGeneral}) and using Equation
(\ref{Eq:AlternatePropagator}) \bea {\cal C} &=& \int dt_s
\expect{\tbE_{M_S}^\dagger(x_s)\tbE_{M_S}(x_s')} \nne \int dt_s
d^3x_s d^3x_s' M_{s}(\bx_s) M_{s}^*(\bx_s') \nnt
\delta(z_s-z_0)\delta(z_s'-z_0)
\expect{\tbE_{s}^\dagger(x_s)\tbE_{s}(x_s') }\nne
\frac{|\tbE_{Mp}|^2\omega_s^4}{|\BVConst_{z,s}|^2}\int d^3x d^3x'
M_{s}(\bx) g(\bx) M_{p}^*(\bx) \nnt
\expect{\tbE_{0i}(x)\tbE_{0i}^\dagger(x') }M_{s}^*(\bx') g(\bx')
M_{p}(\bx')\nne \frac{2\hbar  \omega_i \omega_s^2 d^2}{c^3 n_s^2
n_i \varepsilon_0}{|\tbE_{Mp}|^2}  \int d^4 x'' \delta(z''-z_0)
\nnt \left|\BVConst_{i,z} \int d^3x G_i(x'';x) M_{s}^*(\bx) m(\bx)
M_{p}(\bx)\right|^2\eea
%We expand the pump fields using the mode
%shape $M_p$ and the idler fields using Equation
%(\ref{Eq:AlternatePropagator}) to get \bea {\cal C} &=& \frac{2
%\hbar \omega_i^3 n_i  \omega_s^4}{c^3 \varepsilon_0} |\tbE_{0p}|^2
%\int dt_s'  \int d^4 x_i \delta(z_i-z_f) \nnt \int d^4x'
%\Green_{s}^*(t_s',M_S;x')   g(x') M_p^*(\bx') \nnt
%\Green_{i}^*(x_i;x')\nnt \int d^4x'' \Green_{s}(t_s'',M_S;x'')
%g(x'') M_p(\bx'') \nnt \Green_{i}(x_i;x'') \eea and we use the
%translational symmetry to change $dt_i dt''$ to $dt_s'' dt''$ and
%do the integrals over $t_s',t_s''$ to turn the signal Green
%functions into mode functions, so that  \bea {\cal C} &=&
%  \frac{ \hbar c \omega_i \omega_s^2}{8 n_s^2 n_i \varepsilon_0}
%|\tbE_{0p}|^2  \int d^4 x_i  \delta(z_i-z_f) \delta(t_i-t_f)  \nnt \left|\BVConst_{i,z} \int
%d^4x' \Green_{i}(x_i;x') M_{s}(\bx')g(\bx') M_p(\bx') \right|^2 . \eea
so that
% \bea \label{Eq:SinglesRateSignal}W^{(1)} &=&  \frac{n_s c
%\varepsilon_0 }{2 \hbar \omega_s} \frac{ \hbar c \omega_i
%\omega_s^2}{8 n_s^2 n_i \varepsilon_0} |\tbE_{0p}|^2 I_{DFG}^{(s)}
%\Gamma_{{\rm eff},s} \nne \frac{c^2 \omega_i\omega_s}{16 n_i  n_s}
%|\tbE_{0p}|^2  I_{DFG}^{(s)} \Gamma_{{\rm eff},s} \nne \frac{c^2
%\omega_i\omega_s}{16 n_i  n_s} \frac{P_p}{2 n_p c \varepsilon_0}
%I_{DFG}^{(s)}  \Gamma_{{\rm eff},s} \nne  \frac{\omega_s}{4
%\omega_i}\Gamma_{{\rm eff},s} {P_p} Q_{DFG}^{(s)} \eea
\bea \label{Eq:SinglesRateSignal}W^{(1)} &=& \frac{ c \omega_i
\omega_s}{32  n_p n_s n_i \varepsilon_0} {P_p} |I_{DFG}^{(s)}|^2
\Gamma_{{\rm eff},s} \nne \frac{\omega_s}{4 \omega_i}\Gamma_{{\rm
eff},s} {P_p} Q_{DFG}^{(s)} \eea

\subsection{conditional efficiency}

The conditional efficiency for the idler (probability of
collecting the idler, given that the signal was collected) is \bea
\label{Eq:PDCSignalEfficiency} \eta_s &\equiv&
\frac{W^{(2)}}{{W_s^{(1)}}} = \frac{\Gamma_{\rm
eff}}{\Gamma_{s}}\frac{|I_{SFG}|^2}{|I_{DFG}^{(s)}|^2} \eea
%\bea
%\label{Eq:PDCSignalEfficiency} \eta_s &\equiv&
%\frac{W^{(2)}}{{W_s^{(1)}}} \nne \frac{\omega_i^2
%}{\omega_p^2}\frac{\Gamma_{\rm
%eff}}{\Gamma_{s}}\frac{Q_{SFG}}{{Q_{DFG}^{(s)}}} \nne
%\frac{\omega_i^2}{\omega_p^2}\frac{\Gamma_{\rm
%eff}}{\Gamma_{s}}\frac{ c \omega_p^2 }{8 \varepsilon_0 n_p n_s n_i
%}\frac{8 \varepsilon_0 n_s n_i n_p}{c \omega_i^2}
%\frac{|I_{SFG}|^2}{|I_{DFG}^{(s)}|^2} \nne \frac{\Gamma_{\rm
%eff}}{\Gamma_{s}}\frac{|I_{SFG}|^2}{|I_{DFG}^{(s)}|^2} \eea

\subsection{degenerate processes}

Up to this point, we have discussed only non-degenerate processes,
i.e., those in which the signal and idler fields are distinct and
do not interfere.  This is always the case for type-II
down-conversion, and will be the case for type-I down-conversion
if the frequencies and/or directions of propagation are
significantly different.  We now consider degenerate processes, in
which there is only one down-converted field (signal).

The above discussion is modified only slightly.  The signal and
pump evolve by \bea {\cal D}_p \tbE_p &=& \frac{1}{2}\omega_p^2 g
\tbE_s \tbE_s \nn {\cal D}_s \tbE_s &=& \omega_s^2 g \tbE_p
\tbE_s^\dagger.\eea

\subsection{second harmonic generation}

The calculation of second-harmonic generation (SHG) proceeds
exactly as in sum-frequency generation, except for the factor of
one half and with all ``idler'' variables replaced by ``signal''
variables. Thus we find \bea P_p = P^2_s \Eff_{SHG} \eea where
\bea \label{Eq:SHGEfficiency} \Eff_{SHG} &=& \frac{ \omega_p^2 d^2
}{2 c^3\varepsilon_0 n_p n_s^2 }|I_{SHG}|^2\eea and \bea
\label{Eq:SHGOverlap} I_{SHG} & \equiv & \int d^3x
M_p^*(\bx)m(\bx) M_s(\bx)M_s(\bx) .\eea

\subsection{average parametric gain}

The other classical process of interest is parametric
amplification of the signal by the pump.  The first-order solution
for the signal field is \bea \tbE_s & = & \tbE_{0s} +  \omega_s^2
\int d^4x' \Green_{s}(x;x')  \nnt g(x')
\tbE_{0p}(x')\tbE_{0s}^\dagger(x')\nnequiv \tbE_{0s} + \tbE_{1s} .
\eea  The signal power at the output is \bea P_s &=& 2 n_s c
\varepsilon_0 \int d^3 x_s \delta(z_s-z_0)|\tbE_s(x_s)|^2  \nne 2
n_s c \varepsilon_0 \int d^3 x_s \delta(z_s-z_0)   \left(
|\tbE_{0,s}(x_s)|^2\right. \nnp 2
Re[\tbE_{0,s}(x_s)\tbE_{1,s}^*(x_s)] + \left.
|\tbE_{1,s}(x_s)|^2\right).\eea  The first term is the input
signal power $P_{0,s}$, the second term depends on the relative
phase $\phi_p - 2 \phi_s$, and the last term is the
phase-independent contribution to the gain, an experimentally
accessible quantity. We have \bea \overline{\delta P}
&\equiv&\expect{P_s - P_{0,s}}_{\phi_s} \nne 2 n_s c \varepsilon_0
\int \int d^3 x_s \delta(z_s-z_0) |\tbE_{1,s}(x_s)|^2 \nne \frac{8
\omega_s^2  \varepsilon_0 d^2}{cn_s} |\tbE_{0s}|^2  |\tbE_{p}|^2
\int d^4 x_s \delta(z_s-z_0) \nnt \left|\BVConst_{z,s} \int d^3x'
\Green_{s}(x_s;x') \right. \nnt \left. m(\bx')M_p(\bx')M_s^*(\bx')
\right|^2 \nnequiv  P_{0s} P_{p}\frac{2 \omega_s^2 d^2}{c^3 n_s^2
n_p \varepsilon_0} |I_{APG}|^2 \nnequiv P_{0s} P_{p} Q_{APG}.\eea

\subsection{degenerate PDC}

Next we consider the process of degenerate parametric
down-conversion, for which \bea \tbE_s & = & \tbE_{0s} +
\omega_s^2 \int d^4x' \Green_{s}(x;x')  \nnt g(x')
\tbE_{0p}(x')\tbE_{0s}^\dagger(x'). \eea  We find the correlation
function \bea \expect{\tbE_{s}(x_s)\tbE_{s}(x_s')} &=& \omega_s^2
\int d^4 x'' \Green_s(x_s,x'')
\nnt\expect{\tbE_{0,s}(x_s')\tbE^\dagger_{0,s}(x'')} g(x'')\nnt
\tbE_{0,p}(x'')\eea at which point it is clear that the only
difference from the non-degenerate case of Eq.
(\ref{Eq:NDPDCCorrFun})will be the replacement of idler variables
with signal variables. We find \bea W^{(2)} &=&
\label{Eq:DegPDCPairRate} \Gamma_{\rm eff}
\frac{\omega_s^2}{4\omega_p^2 }P_p  \Eff_{SHG} = \frac{\Gamma_{\rm
eff}}{16}P_p  \Eff_{SHG}.\eea

%We note that this differs by a factor of four from the
%non-degenerate efficiency in Equation (\ref{Eq:PDCPairRate}).  The
%difference lies in the definitions of $\Eff_{SFG}$ and
%$\Eff_{SHG}$.  For example: consider SFG and SHG with the same
%total input and output powers. If the SFG input power is equally
%divided between signal and idler, $\Eff_{SFG}=4 \Eff_{SHG}$.

\subsection{Singles rates (degenerate)}

As before, we can find the rate of detection of singles in the
mode $M_S$ by equation (\ref{Eq:SinglesRateGeneral}) and \bea
{\cal C} &=& \int dt_s' \expect{ \tbE_{M_S}^\dagger(t_s')
\tbE_{M_S}(t_s'')} \nne \int dt_s' d^3x_s' d^3x_s' M_{s}(\bx_s')
M_{s}^*(\bx_s'') \nnt \delta(z_s'-z_0)\delta(z_s''-z_0)
\expect{\tbE_{s}^\dagger(x_s')\tbE_{s}(x_s'') }\nne
\frac{|\tbE_{Mp}|^2\omega_s^4}{|\BVConst_{z,s}|^2}\int d^3x'
d^3x'' M_{s}(\bx') g(\bx') M_{p}^*(\bx') \nnt
\expect{\tbE_{0i}(x')\tbE_{0i}^\dagger(x'') }M_{s}^*(\bx'')
g(\bx'') M_{p}(\bx'') \nne \frac{2 \hbar  \omega_s^3 d^2}{ c^3
n_s^3 \varepsilon_0}{|\tbE_{Mp}|^2}  \int d^4 x'' \delta(z''-z_0)
\nnt \left|\BVConst_{s,z} \int d^3x G_s(x'';x) M_{s}^*(\bx) m(\bx)
M_{p}(\bx)\right|^2 .\eea
%We expand the pump fields using the mode shape $M_p$ and the
%signal fields using Equation (\ref{Eq:AlternatePropagator}) to get
%\bea {\cal C} &=& \frac{8 \hbar n_s  \omega_s^7}{c^3
%\varepsilon_0} |\tbE_{0p}|^2 \int dt_s' \int d^4 x_s'''
%\delta(z_s'''-z_f) \nnt \int d^4x' \Green_{s}^*(t_s',M_S;x') g(x')
%M_p^*(\bx') \nnt \Green_{s}^*(x_s''';x')\nnt \int d^4x''
%\Green_{s}(t_s'',M_S;x'') g(x'') M_p(\bx'') \nnt
%\Green_{s}(x_s''';x'') \eea and we use the translational symmetry
%to change $dt_f dt''$ to $dt_s'' dt''$ and do the integrals over
%$t_s',t_s''$ to turn the Green functions into mode functions, so
%that  \bea {\cal C} &=& \frac{\hbar c \omega_s^3}{2 n_s^3
%\varepsilon_0} |\tbE_{0p}|^2 \int d^4 x_s''' \delta(z_s'''-z_f)
%\delta(t_s'''-t_f)   \nnt \left|\BVConst_{z,s} \int d^4x'
%\Green_{i}(x_s''';x') M_{s}(x')
%%\right. \nnt \left. \rule{0pt}{14pt}
%g(x') M_p(\bx') \right|^2 .
%\eea
The singles rate is thus \bea
\label{Eq:DegSinglesRateSignal}W^{(1)}_s &=&
%\frac{n_s c
%\varepsilon_0 }{2 \hbar \omega_s} \frac{\hbar c \omega_s^3}{2
%n_s^3 \varepsilon_0} |\tbE_{0p}|^2  I_{APG}^{(s)} \Gamma_{{\rm
%eff},s} \nne \frac{c^2 \omega_s^2 }{4  n_s^2} |\tbE_{0p}|^2
%I_{APG}^{(s)}  \Gamma_{{\rm eff},s} \nne \frac{c^2 \omega_s^2 }{4
%n_s^2} \frac{P_p}{2 n_p c \varepsilon_0} I_{APG}^{(s)}
%\Gamma_{{\rm eff},s} \nne
\frac{1}{4}\Gamma_{{\rm eff},s} {P_p}
Q_{APG}^{(s)}.\eea

\subsection{Conditional efficiency (degenerate)}

The conditional efficiency is \bea
\label{Eq:DegPDCSignalEfficiency} \eta_s &\equiv&
\frac{W^{(2)}}{{W_s^{(1)}}} =
%\nne \frac{4\omega_s^2
%}{\omega_p^2}\frac{\Gamma_{\rm eff}}{\Gamma_{{\rm
%eff},s}}\frac{Q_{SHG}}{{Q_{APG}^{(s)}}} \nne
%\frac{4\omega_i^2}{\omega_p^2}\frac{\Gamma_{\rm eff}}{\Gamma_{{\rm
%eff},s}}\frac{ c \omega_p^2 }{8 \varepsilon_0 n_p n_s^2 }\frac{2
%\varepsilon_0 n_s^2 n_p}{c \omega_s^2}
%\frac{|I_{SHG}|^2}{|I_{APG}^{(s)}|^2} \nne
\frac{\Gamma_{\rm eff}}{\Gamma_{{\rm
eff},s}}\frac{|I_{SHG}|^2}{|I_{APG}^{(s)}|^2} \eea
\newline

\section{Example calculations}

\newcommand{\geomfactor}{\Upsilon}

 We now illustrate the preceding, general results with a few special cases. We first
calculate the overlap integral for co-propagating gaussian beams.
This allows us to 1) compare our results to the classical results
of Boyd and Kleinman \cite{Boyd:1968}, 2) predict absolute
brightness for an important geometry, type-II co-linear
down-conversion in quasi-phase-matched material.  Also, we compare
to a recent calculation of absolute brightness for a specific
geometry by Ling, et al. \cite{Ling:2008}.

We consider collinear, frequency-degenerate type-II PDC with
circular gaussian beams for signal, idler and pump.   We take mode
shape functions \be M_{m}(\bx) = \sqrt{\frac{k_m
z_R}{\pi}}\frac{1}{q} e^{i k_m z} e^{i k_m \frac{r^2}{2 q}}.\ee
where $m \in \{s,i,p\}$, $r$ is the radial component of $\bx$, and
$q \equiv z-i z_R$ where $z_R$ is the Rayleigh range, assumed
equal for all beams.  We assume a periodically-poled material in
which $\chitwo(z)$ alternates with period $2 \pi/Q$ so and we
approximate $m(\bx) \approx \exp[i Q z]\deff/d $.  From Equation
(\ref{Eq:SingleModeSFGBrightness}) we find \bea I_{SFG} &=&
\sqrt{\frac{k_p k_s k_i z_R^3}{\pi^3}} \int_{-L/2}^{L/2} dz'
\frac{e^{-i \Delta k z'}}{q|q|^2 } \nnt \int 2 \pi r' dr' e^{-i
\left( \frac{k_p}{q^*} - \frac{k_s+k_i}{q}\right) \frac{r'^2}{2} }
\nne \frac{2 i}{k_+}\sqrt{\frac{k_p k_s k_i z_R^3}{\pi}} \nnt
\int_{-L/2}^{L/2} dz' \frac{e^{-i \Delta k z'}}{(z'-iz_R)\left(R_k
z'+iz_R\right)} \eea where $\Delta k \equiv k_p - k_s - k_i - Q$
and $R_k \equiv k_-/k_+$ and $k_\pm \equiv k_p \pm (k_s + k_i)$.
In terms of the dimensionless variables $\kappa \equiv \Delta k
L$, $\zeta \equiv z'/L$, $\zeta_R \equiv z_R/L$ we find \bea
I_{SFG} &=& \frac{2i}{k_+} \sqrt{\pi k_p k_s k_i z_R } \geomfactor
%(\zeta_R,\kappa,R_k)
\eea where \bea \geomfactor &\equiv & \frac{\zeta_R}{2
\pi}\int_{-1/2}^{1/2} d\zeta \frac{e^{-i \kappa
\zeta}}{(\zeta-i\zeta_R)\left(R_k \zeta+i\zeta_R\right)}. \eea

From Equation (\ref{Eq:SingleModeSFGEff}) the upconversion
efficiency is then
%\be \Eff_{SFG} = \frac{ 2 \omega_p^2 d^2}{c^3
%\varepsilon_0 n_p n_s n_i }  \frac{4\pi k_p k_s k_i z_R }{k_+^2}
%|\geomfactor|^2 \ee
\be \Eff_{SFG} = \frac{ 8 \pi \omega_p^2 }{
c^3 \varepsilon_0 n_p n_s n_i } \frac{k_p k_s k_i }{k_+^2} z_R
\deff^2 |\geomfactor|^2. \ee

\subsection{Boyd and Klienman, 1968}

With this expression we can compare our results to those of Boyd
and Kleinman \cite{Boyd:1968} for the case of second-harmonic
generation.  As that calculation does not include quasi-phase
matching, we take $Q=0$, and then for any reasonable
phase-matching we have $n_p \approx n_s$, $k_p \approx 2 k_s$ and
thus $k_-=R_k=0$, $k_+ \approx 2 k_p$.  We note that for $R_k =
0$, $\geomfactor$ becomes equal to the function $H$ of Boyd and
Kleinman for zero absorption and walk-off angle.  We find \be P_p
= \frac{4 \pi k_s \omega_s^2}{c^3 \varepsilon_0 n_s^2 n_p} z_R d^2
|\geomfactor|^2 P_s^2.\ee

Boyd and Kleinman find in Equations (2.16),(2.17) and (2.20) to
(2.24) \be P_{2} = \frac{128 \pi^2 \omega_1^2}{c^3 n_1^2 n_2} d^2
P_{1}^2 L k_1 \frac{2\pi^2 z_R}{L} |H|^2\ee or \be P_{2} =
\frac{256 \pi^4 k_1\omega_1^2}{c^3 n_1^2 n_2}z_R d^2 |H|^2
P_{1}^2.\ee   When converting this expression to MKS units, $d^2
\rightarrow d^2/64 \pi^3 \varepsilon_0$, and we see that the two
calculations agree.

\subsection{Type-II collinear brightness}

Next we make a numerical calculation for frequency-degenerate
type-II SPDC, a geometry of current interest for generation of
entangled pairs, for example.

The integral $\geomfactor$ must be evaluated numerically.  For a 1
cm crystal of PPKTP and a vacuum wavelength $\lambda_s = \lambda_i
= 800$ nm we have $(n_s,n_i,n_p) = (1.844, 1.757, 1.964)$ and
$d_{\rm eff} = 2.4$ pm/V so that $R_k = 0.04$ and the maximum of
$(z_R/L)|\geomfactor|^2\approx 0.054$ occurs at $\kappa \approx
-3.0, \zeta_R \approx 0.18$.  We find $\Eff_{SFG} = 2.0\times
10^{-3}$ W$^{-1}$.  Used as a photon-pair source, this same
crystal and geometry would yield by Equation
(\ref{Eq:PDCPairRate}) \be W^{(2)} = \Gamma_{\rm eff} P_p
\frac{\Eff_{SFG}^{}}{16}  \ee or a pair generation efficiency of
$\Eff_{SFG}/16 = 0.8 $ pairs
%per second per mW of pump power per GHz of filter bandwidth.
(s mW MHz)$^{-1}$.  Note that $\Gamma_{\rm eff}$ is the filter
bandwidth in angular frequency.

\subsection{Ling, Lamas-Linares, and Kurtsiefer, 2008}
\label{Sec:Ling}

Recently, Ling {\em et al.} calculated the absolute emission rate
into gaussian modes in the thin-crystal limit of
(non-periodically-poled) nonlinear material \cite{Ling:2008}. They
arrive to a down-conversion spectral brightness of \be
\label{Eq:LingRate} \frac{dR(\omega_s)}{d\omega_s} =
\left(\frac{\deff \alpha_s \alpha_i E_p^0 \Phi(\Delta k)}{c}
\right)^2 \frac{\omega_s\omega_i}{2 \pi n_s n_i} \ee where $R$ is
the pair collection rate and \be \Phi(\Delta k) \equiv \int dz
\int dy\, dx\, e^{i \Delta \bk \cdot \br }
U_p(\br)U_s(\br)U_i(\br).\ee Here $U_m$ describe the mode shapes
of the form $U_m(\br) = e^{i k_m z} e^{-(x^2 + y^2)/W_m^2}$ and
$\alpha_m = \sqrt{2/\pi W_m^2}$ are normalization constants. The
field $E_p^0$ is defined such that $|E_p^0|^2 = 2\alpha_p^2
P_p/\varepsilon_0 n_p c$ where $P_p$ is the pump power, giving \be
\frac{dR(\omega_s)}{d\omega_s} = \frac{\omega_s\omega_i d^2}{\pi
c^3\varepsilon_0 n_p n_s n_i} P_p \frac{}{
}|\alpha_p\alpha_s\alpha_i \Phi(\Delta k)|^2 \ee Assuming the
output is collected with narrow-band filters of transmission $T_s
(\omega_s),T_i(\omega_i)$ for signal and idler, respectively, the
integrated rate is \be R = \frac{dR(\omega_s)}{d\omega_s} \int
d\Omega T_s(\omega_p/2+\Omega) T_i(\omega_p/2-\Omega) \ee where we
have assumed $dR(\omega_s)/d\omega_s$ constant over the width of
the filters. For comparison, using Eqs (\ref{Eq:SingleModeSFGEff})
and (\ref{Eq:PDCPairRate}), we find \be W^{(2)} = \frac{\omega_s
\omega_i d^2 }{2c^3 \varepsilon_0 n_p n_s n_i} P_p |I_{SFG}|^2
\Gamma_{\rm eff}.\ee Then with Equation (\ref{Eq:GammaArea}) and
noting that in the thin crystal limit $|I_{SFG}| =
|\alpha_p\alpha_s\alpha_i \Phi(\Delta k)|$, we see that the two
results are identical.

%rectangular filter \be T(\Omega) = \left\{
%\begin{array}{cl}1 & |\Omega| \le \delta/2 \\ 0 & |\Omega| >
%\delta/2 \end{array} \right. \ee has $\Gamma_{\rm eff} = 2
%\delta/\pi$.  Thus the two calculations give the same

%273000 pairs (s mW nm)-1, fabrizi
%
% Or in terms of the dimensionless
%variables $\tau \equiv z'/z_R$, $\sigma \equiv \Delta k z_R$,
%$\tau \equiv z'/z_R$, $\xi \equiv L/2 z_R$, \bea I_{DC} &=&
%i\sqrt{\frac{k_p k_s k_i }{\pi^3}}\frac{d_{\rm eff}}{c^2}
%z_R^{1/2} \nnt {}\int_{-\xi}^{\xi} d\tau \frac{e^{i \sigma
%\tau}}{(1+i\tau)\left(1 + i R_k \tau \right)} \eea

%\bea I_{DC} &=& \sqrt{\frac{k_p k_s k_i z_R^3}{\pi^3}}\frac{d_{\rm
%eff}}{c^2} \int_{-L/2}^{L/2} dz' \frac{e^{i \Delta k z'}}{(q^*)^2
%q} \nnt \int 2 \pi r' dr' e^{-i \left( \frac{k_p}{q} -
%\frac{k_s+k_i}{q^*}\right) \frac{r'^2}{2} } \nne -i\sqrt{\frac{k_p
%k_s k_i z_R^3}{\pi^3}}\frac{d_{\rm eff}}{c^2} \nnt
%\int_{-L/2}^{L/2} dz' \frac{e^{i \Delta k z'}}{(z'-iz_R)\left(R_k
%z'-iz_R\right)} \eea where $\Delta k \equiv k_p - k_s - k_i - Q$
%and $R_k \equiv ({k_p - k_s - k_i})/({k_p + k_s + k_i})$.  Or in
%terms of the dimensionless $\tau \equiv z'/z_R$, $\sigma \equiv
%\Delta k z_R$, $\tau \equiv z'/z_R$, $\xi \equiv L/2 z_R$, \bea
%I_{DC} &=& i\sqrt{\frac{k_p k_s k_i }{\pi^3}}\frac{d_{\rm
%eff}}{c^2} z_R^{1/2} \nnt {}\int_{-\xi}^{\xi} d\tau \frac{e^{i
%\sigma \tau}}{(1+i\tau)\left(1 + i R_k \tau \right)} \eea

\section{Conclusions}
\label{sec:Conclusions}

Using the approach of coupled wave-equations, familiar from
nonlinear optics, we have calculated the absolute brightness and
temporal correlations of spontaneous parametric down-conversion in
the narrow-band regime.  The results are obtained with a Green
function method and are generally valid within the paraxial
regime.  We find that efficiencies of SFG and SPDC can be
expressed in terms of mode overlap integrals, and are proportional
for corresponding geometries. Also, we find pair time correlations
in terms of signal and idler filter impulse response functions.
Results for both degenerate and non-degenerate SPDC are found.
Comparison to classical calculations by Boyd and Kleinman, and to
a recent calculation by Ling et al. show the connection to
classical nonlinear optics and ``golden rule''-style brightness
calculations, while considerably generalizing the latter.  We
expect these results to be important both for designing SPDC
sources, as the results of well-known classical calculations can
be used, and for building and optimizing such sources.

% We have considered nonlinear optical
%processes with quantum fields in the paraxial and narrow-band
%regime.  Coupled paraxial wave equations describe the three-wave
%mixing processes and Green functions describe the spatio-temporal
%evolution of the fields. We find expressions for the
%time-correlations of the output in terms of temporal filters, and
%show that the efficiency of single-mode processes are given in
%terms of mode-overlap integrals. Furthermore, the efficiency of
%spontaneous parametric down-conversion can be found from the
%efficiency for sum-frequency generation under very general
%conditions. The photon collection efficiency can be found by
%comparing the efficiencies of SFG and difference-frequency
%generation.  Similar relationships are found for degenerate signal
%and idler fields and corresponding classical three-wave mixing
%processes.  We expect these results to be important both for
%designing SPDC sources, as the results of well-known classical
%calculations can be used, and for building and optimizing such
%sources.
%

\appendix{}

\section{alternate propagator}

We can use Equations (\ref{Eq:BackFromTheFuture}) and
(\ref{Eq:EqualTimeCommutator}) to express the propagator as \bea
\expect{\tbE_{}(x)\tbE_{}^\dagger(x') }&=& |\BVConst_{t}|^2 \int
d^4 x'' d^4 x''' \delta(t''-t_f) \nnt \delta(t'''-t_f)
\expect{\tbE_{}(x'')\tbE_{}^\dagger(x''') } \nnt \Green^*(x'';x)
 \Green(x''';x').
% We
%note that $ \expect{\tbE_{}(t_f,\bx'')\tbE_{}^\dagger(t_f,\bx''')
%} = A^2_{\gamma} \delta^3(\bx''-\bx''')$
%%$\expect{\tbE_{}(t_f,\bx'')\tbE_{}^\dagger(t_f,\bx''') } =
%%\delta^3(\bx''-\bx''')\hbar\omega v_{g}/2n c\varepsilon_0$
%to get
\nne |A_{\gamma} \BVConst_{t}|^2 \int d^4 x''  \delta(t''-t_f)
\nnt \Green^*(x'';x)
 \Green(x'';x') \eea Noting that $\int d^4 x'' \delta(t''-t_f)  \Green^*(x'';x)
 \Green(x'';x') = v_g
\int d^4 x'' \delta(z''-z_0)  \Green^*(x'';x)
 \Green(x'';x') $ we find
\bea \label{Eq:AlternatePropagator}
\expect{\tbE_{}(x)\tbE_{}^\dagger(x') }&=&
  \frac{2\hbar n \omega^3 }{c^3 \varepsilon_0}
\int d^4 x'' \delta(z''-z_0) \nnt  \Green^*(x'';x) \Green(x'';x')
\eea
%\bea \label{Eq:AlternatePropagator}
%\expect{\tbE_{}(x)\tbE_{}^\dagger(x') }&=& \frac{\hbar \omega }{2
%n c \varepsilon_0} \frac{4 n^2 \omega^2}{c^2v_g^2}v_g^2 \int d^4
%x'' \nnt \delta(z''-z_f)  \Green^*(x'';x) \nnt
% \Green(x'';x')
%  \nne
%  \frac{2\hbar n \omega^3 }{c^3 \varepsilon_0}
%\int d^4 x'' \delta(z''-z_f) \nnt  \Green^*(x'';x) \Green(x'';x')
%\eea

\section{Lorentzian filter}

A common filter has a Lorentzian transfer function and an
exponential impulse response \be F(\tau) = \frac{\Gamma}{2}
\theta(\tau) \exp[-\Gamma \tau /2 ].\ee  The spectral transmission
is $T(\Omega) = \Gamma^2/(\Gamma^2 + 4 \Omega^2)$, i.e., unit
transmission for constant $\tbE$, a full-width at half-maximum of
$\Delta \Omega_{\rm FWHM} = \Gamma$ and an area $\int d\Omega\,
T(\Omega) = \pi\Gamma /2$.  If we put a filter of this sort in
each arm, the output has \bea f(t_s-t_i) &= &
\frac{\Gamma_s\Gamma_i}{4}\int dt' \theta(t_s-t') \theta(t_i-t')
\nnt \exp[-\Gamma_i(t_i-t')/2]\nnt \exp[-\Gamma_s(t_s-t')/2] \eea
or \bea f(\tau) &=& \frac{\Gamma_s\Gamma_i}{2(\Gamma_s +
\Gamma_i)} \left\{
\begin{array}{ll}\exp[-\Gamma_s\tau/2] &
\tau > 0 \\
\exp[\Gamma_i\tau/2] & \tau < 0  \end{array} \right. \nn \eea The
effective bandwidth is \be \Gamma_{\rm eff} = 4 \int d\tau
|f(\tau)|^2= \frac{\Gamma_s\Gamma_i}{\Gamma_s+\Gamma_i} \ee

It is worth noting that in the limit $\Gamma_s \rightarrow \infty$
(the limit of a broad-band filter in the signal beam, or in
practical terms, not having a filter there at all), filter becomes
\bea f(t_s-t_i)  &= & \frac{\Gamma_i}{2} \left\{
\begin{array}{rl}0 &
t_i < t_s \\
\exp[-\Gamma_i(t_i-t_s)/2] & t_i > t_s \end{array}\right. \nn\eea
That is, the idler photon will always arrive later, and with a
distribution (after the signal arrival) that is precisely the
transfer function of the idler-beam filter. Another interesting
limit is for matched filters, $\Gamma_s = \Gamma_i = \Gamma$. Then
we find \bea f(t_s-t_i) &= &
\frac{\Gamma}{4}\exp[-\Gamma|t_s-t_i|/2] .\eea

%\subsubsection{intrinsic inefficiency of filtering both sides}

Note that for $\Gamma_s \rightarrow \infty$, the detection rate is
$|{\cal A}|^2 \Gamma_i /{4} $, i.e., proportional to the idler
filter bandwidth $\Gamma_i$.  The reverse, $s \leftrightarrow i$
is also true, of course.  From this we can get an idea of the
conditional efficiency:  The rate for filtered signal with {any}
idler is proportional to \be \Gamma_s \ge \frac{\Gamma_i
\Gamma_s}{\Gamma_s + \Gamma_i}.\ee  For example, putting matched
filters $\Gamma_s = \Gamma_i = \Gamma$ will give a rate
proportional to $\Gamma_i \Gamma_s/({\Gamma_s + \Gamma_i})$ i.e.,
half of the rate without the idler filter. This indicates that, of
the signal photons that pass the the signal filter, half of their
``twin'' idler photons do not pass the idler filter.

\subsection*{Acknowledgements}
We thank A. Cer\`{e}, F. Wolfgramm, G. Molina, A. Haase, and N.
Piro for helpful discussions.  This work was supported by the
Spanish MEC under the ILUMA project (Ref. FIS2008-01051), the
Consolider-Ingenio 2010 Project
\textquotedblleft{}QOIT\textquotedblright{} and by Marie Curie RTN
``EMALI.''

%\cite{Klyshko:1969,Klyshko:1969a}

\bibliographystyle{apsrev}
\bibliography{NBDC8}

\vspace{\fill}

%
%\begin{thebibliography}
%references here.
%\end{thebibliography}

\end{document}

%% file: NewCommands.tex
\newcommand{\be}{\begin{equation}}
\newcommand{\ee}{\end{equation}}
\newcommand{\bea}{\begin{eqnarray}}
\newcommand{\eea}{\end{eqnarray}}
\newcommand{\sinc}{\mbox{\rm sinc}}
\newcommand{\ba}{{\bf a}}
\newcommand{\bb}{{\bf b}}
\newcommand{\bd}{{\bf d}}
\newcommand{\bk}{{\bf k}}
\newcommand{\bq}{{\bf q}}
\newcommand{\bJ}{{\bf J}}
\newcommand{\bj}{{\bf j}}
\newcommand{\bE}{{\bf E}}
\newcommand{\bx}{{\bf x}}
\newcommand{\br}{{\bf r}}
\newcommand{\bH}{{\bf H}}
\newcommand{\bB}{{\bf B}}
\newcommand{\bP}{{\bf P}}
\newcommand{\bC}{{\bf C}}
\newcommand{\bD}{{\bf D}}
\newcommand{\bv}{{\bf v}}
\newcommand{\brho}{{\bf \rho}}
\newcommand{\balpha}{\stackrel{\leftrightarrow}{\alpha}}
\newcommand{\twovector}[2]{\left[\begin{array}{c}#1\\#2\end{array}\right]}
\newcommand{\twomatrix}[4]{\left[\begin{array}{cc}#1&#2\\#3&#4\end{array}\right]}
\newcommand{\threevector}[3]{\left[\begin{array}{c}#1\\#2\\#3\end{array}\right]}
\newcommand{\threematrix}[9]{
\left[\begin{array}{ccc}#1&#2&#3\\#4&#5&#6\\#7&#8&#9\end{array}\right]}
\newcommand{\braket}[2]{\left<#1|#2\right>}
\newcommand{\ket}[1]{\left|#1\right>}
\newcommand{\bra}[1]{\left<#1\right|}
\newcommand{\Schrodinger}{Schr\"{o}dinger~}
\newcommand{\FigWidth}{15 cm}
\newcommand{\expect}[1]{\left<#1\right>}
\newcommand{\adag}{{a^\dagger}}
\newcommand{\Tr}{{\rm Tr}}
\newcommand{\chione}{\chi^{(1)}}
\newcommand{\chitwo}{\chi^{(2)}}
\newcommand{\chithr}{\chi^{(3)}}
\newcommand{\deff}{d_{\rm eff}}
\newcommand{\bee}{{\bf e}}
\newcommand{\bA}{{\bf A}}
\newcommand{\bnab}{{\bf \nabla}}
\newcommand{\epzero}{\varepsilon_{0}}
\newcommand{\exercise}{\hspace{-.25in}\Large}
\newcommand{\hH}{\hat H}
\newcommand{\hn}{\hat n}
\newcommand{\ha}{\hat a}
\newcommand{\hd}{\hat d}
\newcommand{\hx}{\hat x}
\newcommand{\hp}{\hat p}
\newcommand{\hadagger}{\hat a^{\dagger}}
\newcommand{\hb}{\hat b}
\newcommand{\hbdag}{\hat b^{\dagger}}
\newcommand{\hau}{\hat{\underline{a}}}
\newcommand{\haudagger}{\hau^{\dagger}}
\newcommand{\hbA}{\hat b_{A}}
\newcommand{\hbE}{\hat b_{E}}
\newcommand{\hE}{\hat {\bf E}}
\newcommand{\hnbE}{\hat {E}}
\newcommand{\hnbD}{\hat {D}}
\newcommand{\hEp}{\hat {E}^{(+)}}
\newcommand{\hEm}{\hat {E}^{(-)}}
\newcommand{\hX}{\hat {X}}
\newcommand{\tbE}{{\bf\cal E}}
\newcommand{\tbF}{{\bf\cal F}}
\newcommand{\Ep}{E^{(+)}}
\newcommand{\Em}{E^{(-)}}
\newcommand{\tbX}{{\bf\cal X}}
\newcommand{\tP}{{\cal P}}
\newcommand{\tbP}{{\bf\cal P}}
\newcommand{\tE}{{\cal E}}
\newcommand{\tF}{{\cal F}}
\newcommand{\tG}{{\cal G}}
\newcommand{\tX}{{\cal X}}
\newcommand{\cE}{{\cal E}}
\newcommand{\cF}{{\cal F}}
\newcommand{\cD}{{\cal D}}
\newcommand{\cP}{{\cal P}}
\newcommand{\nn}{\nonumber \\ }
\newcommand{\nnn}{\nonumber \\ &  &}
\newcommand{\nne}{\nonumber \\ & = &}
\newcommand{\nnequiv}{\nonumber \\ & \equiv &}
\newcommand{\nnapprox}{\nonumber \\ & \approx &}
\newcommand{\nnpropto}{\nonumber \\ & \propto &}
\newcommand{\nnp}{\nonumber \\ & & +}
\newcommand{\nnm}{\nonumber \\ & & -}
\newcommand{\nnt}{\nonumber \\ & & \times}
\newcommand{\bdp}{\bd^{(+)}}
\newcommand{\bdm}{\bd^{(-)}}
\newcommand{\bEp}{\bE^{(+)}}
\newcommand{\bEm}{\bE^{(-)}}
\newcommand{\bBp}{\bB^{(+)}}
\newcommand{\bBm}{\bB^{(-)}}
\newcommand{\hain}{\ha_{\rm IN}}
\newcommand{\haout}{\ha_{\rm OUT}}
\newcommand{\hadaggerin}{\hadagger_{\rm IN}}
\newcommand{\hadaggerout}{\hadagger_{\rm OUT}}
\newcommand{\hEin}{\hnbE_{\rm IN}}
\newcommand{\hEout}{\hnbE_{\rm OUT}}
\newcommand{\haone}{\ha_1}
\newcommand{\hadaggerone}{\hadagger_1}
\newcommand{\hatwo}{\ha_2}
\newcommand{\hadaggertwo}{\hadagger_2}
\newcommand{\hainone}{\ha_{{\rm IN},1}}
\newcommand{\hadaggerinone}{\hadagger_{{\rm IN},1}}
\newcommand{\haintwo}{\ha_{{\rm IN},2}}
\newcommand{\hadaggerintwo}{\hadagger_{{\rm IN},2}}
\newcommand{\curl}{\nabla \times}
\newcommand{\ve}{\varepsilon}
\newcommand{\Adagger}{A^{\dagger}}
\newcommand{\Bdagger}{B^{\dagger}}
\newcommand{\Cdagger}{C^{\dagger}}
\newcommand{\Ddagger}{D^{\dagger}}
\newcommand{\Lag}{{\cal L}}
\newcommand{\var}{{\rm var}}

\newcommand{\arctanh}{\rm{arctanh}}

\newcommand{\Pe}{P_e}
\newcommand{\Pg}{P_g}
\newcommand{\PeZ}{P_{e0}}
\newcommand{\PgZ}{P_{g0}}